\documentstyle[12pt,aasms4]{article}
\received{}
\accepted{}
\begin{document}

\title{Unpulsed Optical Emission from the Crab Pulsar{\footnote{Based on
observations using the 6m telescope at the Special Astrophysical Observatory of the Russian
Academy of Sciences in Nizhnij Arkhyz, Russia}}}

\author{A. Golden, A. Shearer} 
\affil{Information Technology Centre,
National University of Ireland, Galway, Ireland}
\author {G.M. Beskin}
\affil {Special Astrophysical Observatory, Nizhnij Arkhyz,
Karachai-Cherkessia, Russia}

\begin{abstract}

Based on observations of the Crab pulsar using the TRIFFID high speed 
imaging photometer in the $UBV$ bands using the Special Astrophysical
Observatory's 6m telescope in the Russian Caucasus, we
report the detection of pronounced emission during the so-called `off' 
phase of emission. Following de-extinction, this unpulsed component
of emission is shown to be consistent with a power law with an exponent of 
$\alpha$ = -0.60 $\pm$ 0.37, the uncertainty being dominated by the 
error associated with the independent
CCD photometry used to reference the TRIFFID data.
This suggests a steeper power law form than that reported elsewhere
in the literature for the total integrated spectrum, which is
essentially flat with $\alpha$ $\sim$ 0.1, although the 
difference in this case 
is only significant at the $\leq$ 2$\sigma$ level. 
Deeper reference integrated 
and TRIFFID phase-resolved
photometry in these bands in conjunction with further observations in 
the $UV$ and $R$ region would constrain this fit further. 

\end{abstract}


\section{Introduction}

The Crab pulsar provides one of the best multiwavelength sources of
magnetospheric emission from $\gamma$-rays to the radio regime
and as such, remains the {\it gold standard} as
regards providing definitive empirical datasets with which to
constrain current existing theoretical models of such nonthermal
emission. Throughout this entire frequency range, the pulsar's light curve
retains essentially the same morphology, being traditionally 
divided up into four distinct regions - the two peaks, 
the Bridge of emission between the peaks, and the `off' 
region. This latter component was historically presumed
to originate from the nebula, a reasonable assumption
considering the intense beaming observed
from this object.\\

Optically, the pulsar has been scrutinized ever since its
initial discovery in the radio by \cite{stae68}.
The pulsar is bright enough for effective single-pixel 
high speed photometry, and following its confirmation 
as an optical pulsar by \cite{cocke69}, numerous such 
observations followed (e.g. Wampler et al. 1969, 
Kristian et al. 1970, Cocke and Ferguson 1974,
Groth 1975a, Groth 1975b). These observations typically spanned the 
$BVRI$ wavebands at time resolutions of $\sim$ milliseconds, and 
as absolute reference timing was not possible,
individual light curves per dataset were typically 
co-added in a least-squares fashion.\\

Despite the somewhat restricted data acquisition and 
analytical conditions associated with these observations, 
there was a consensus that the common arrival time of 
all these colored peaks was accurate to within 10$\mu$s, 
there were suggestions of morphological differences 
between the leading and trailing edges of various light curves, 
and that the light curve was strongly polarized as a 
function of rotational phase (Wampler et al. 1969).\\

Subsequent observations by Peterson et al. 1978 
using a 2 dimensional (2-d) image photon counting camera in the 
$UB$ bands suggested that the supposed `off' 
phase of the pulsar's rotational phase was in 
fact consistent with continuing emission from the 
pulsar, indicating the Crab was actually `on' 
for the full rotational cycle. Whilst these 
results at the time were unprecedented, deeper
exposures combined with more rigorous image 
processing algorithms would have yielded more 
accurate estimates of the `off' components
flux yields and overall spectral form.\\ 

In Jones et al. 1981, Smith et al. 1988 and 
Smith et al. 1996, several dedicated phase-resolved 
$V$ \& $UV$ polarimetric observations of the Crab 
pulsar using both ground based single-pixel photometers 
and the $HST$ High Speed Photometer yielded data 
indicating sharp swings in polarisation angle around 
both the peaks, in addition to some form of polarisation 
evolution in the Bridge region. Remarkably, the 
analysis also indicated a large polarisation component 
associated with the traditional `off' phase of emission.
The inference was that, combined with the earlier 
Peterson et al. results, the `off' emission of the 
pulsar was consistent with some form of nonthermal, 
undoubtedly synchrotron related origin. However, the 
single-pixel based nature of these observations limited 
the possibility of accurately resolving the unpulsed 
component's contribution in terms of polarisation, 
which may be expected to contain a substantial nebular 
component.\\

Ideally, one requires a high speed 2-d photometer in 
order to obtain acceptably significant signal to noise (S/N) 
datasets in several wavebands from which one might hope to 
photometrically isolate the various components of a 
phase-resolved light curve.  With such systems, 
the effective photometer sky aperture can be reduced
compared with conventional photometers, and 
effects such as telescope wobble can be entirely 
removed (\cite{shear96}). 
Thus, photons chosen for analysis can be selected in 
software which can place an aperture matched 
(to maximise S/N) to the prevailing seeing 
and background conditions, and then isolate those 
barycentred photons within specifically chosen phase 
regions of the light curve. 
The TRIFFID high speed photometer, previously used 
in the detection of pulsations from both Geminga 
(\cite{shear98}) and PSR B0656+14 (\cite{shear97}) 
is ideal in this regard, as it makes use of a 
MAMA camera. In this communication we 
document the first attempts to photometrically isolate 
the Crab's unpulsed `off' component of emission 
in three color bands.\\

\section{Observations and Analysis}

Observations of the Crab pulsar were made over 5 nights between
the 14th and 19th of January 1996, using the TRIFFID camera mounted
at Prime Focus of the 6m telescope of the
Special Astrophysical Observatory located
in the Russian Caucasus. The primary targets of this observing
run were the Geminga and PSR B0656+14 pulsars, thus the Crab
observations were somewhat limited.
Data was taken in the
$U$, $B$ and $V$ Johnson bands. The plate scale
was 0.22"/pixel at the $MAMA$ photocathode. For all nights
observing the Crab, the atmospheric stability was good 
although there were several transits of high altitude cirrus.
Table \ref{table1} shows a log of the observations. Each individual dataset
was binned to form an integrated image, and from this, reference stars were
chosen as guides for the image processing software. Flat fields were prepared
using deep dome flats co-added with a number of sky flats, taken
immediately after the observations.
Image processing, incorporating a Weiner filter modified shift-and-add 
algorithm (\cite{redf}),
followed, incorporating the derived flat field and correcting
for telescope wobble and gear drift. This yielded full field
images of the inner Crab nebula within which the pulsar and
its stellar companions were registered.\\

\placetable{table1}

For each dataset, all photons within a radius of 50 
pixels of the Crab pulsar were extracted using the image 
processing software, and these time-stamps were
then barycentred using the JPL DE2000 ephemeris.  
A standard epoch folding algorithm was used
to prepare light curves based upon given Jodrell Bank Crab 
Ephemeris (\cite{lyn96}) via folding
modulo the barycentred time series. This yielded both light
curves and a phase-resolved image within a certain specfied
phase range - in this initial case the full cycle. 
The number of phase bins used was 3000, yielding
a bin resolution of $\sim$ 11 $\mu$s.\\

Following this, the s2 ($V$), s1 ($U$) \& w4 ($B$) images
were used respectively as `templates' with which to re-orientate 
the other color datasets geometrically, so as to result in a 
set of identical integrated images for each dataset. 
Total summed errors in this shift-and-rotation technique 
were of order 0.1$\%$ in terms of pixel  units typically. For each 
colour band, each dataset was summed chronologically, 
yielding a total $U$, $B$ and $V$ dataset.\\

\begin{figure}
\plotone{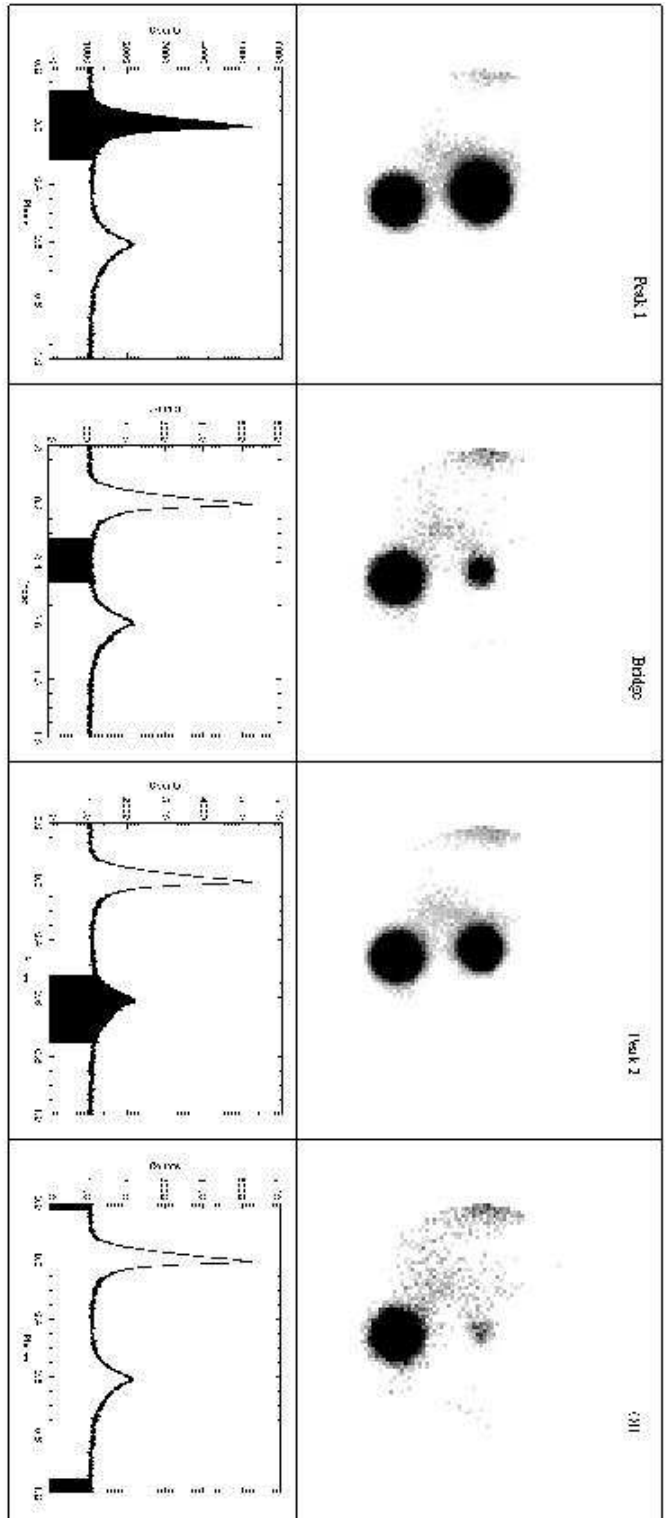}
\caption{Phase-resolved 2-d photometry of the inner Crab nebula
in the $B$ band, taken within a radius of 50 pixels, the pulsar
being the centre star. 
The location of each phase region is indicated in the 
light curve obtained from a radius of 15 pixels from the
Crab centroid, with the accompanying photometric image.\label{fig1}}
\end{figure}

Phase-resolved images obtained based upon the approximate
locations of the four principal morphological regions
previously defined were obtained as shown in Figure 
\ref{fig1}. In this figure, the phase regions defining
the peaks and Bridge of emission match those defined
previously by Eikenberry \& Fazio 1997. It is clear that there is
emission associated with the pulsar during what has been 
conventionally regarded as the `off' phase - as had been 
originally indicated by Peterson et al. With these deep
phase-resolved images, it is possible to apply the full arsenal of image
processing techniques, and thus photometrically characterise
the time-resolved nature of the pulsar's emission particulary
for the `off' region.\\

\placefigure{fig1}

In order to do this, we must satisfactorily isolate the unpulsed
component from the background-removed light curves, in such a way
so that we are satisfied that our denominated phase window
samples what is consistent with an unpulsed component only.
To isolate this `off' region, standard
image processing techniques were used to remove the Crab 
pulsar from each of the full cycle images in the $UBV$ bands.
In effect, one fits an analytical point spread function (PSF)
to the full cycle photometric image, and one then uses this
PSF to firstly derive the flux associated with full cycle
image, and then to derive the fluxes associated with
the other phase resolved images corresponding to the two peaks,
the bridge and the unpulsed component of emission. We
now outline the approach in more detail.
%, and
%then it was a simple matter of determining the background flux within
%a specific aperture.\\ 

For a given color band, the removal of the Crab pulsar from the 
full cycle image was performed via the {\tt daophot} IRAF package, using 
the {\tt psf} task to fit a PSF to
the Crab pulsar stellar point source. This was then used as 
in input to the {\tt allstar} task, which re-fits the PSF 
to the candidate stellar point source - in this first case, the
full cycle Crab image - in order to accurately remove the 
candidate and in so doing, determine both the flux and its
error associated with this procedure.  For the full cycle
image, the removal was performed satisfactorily, as the deep 
exposures in $UBV$ provided good background statistics to the 
required fitting algorithms.\\

Using standard aperture photometry, the resulting Crab-removed 
image was then used to determine the total background flux within
the fixed radius centred on the PSF derived centroid
of the Crab point source. This net background flux was then used
to correct the existing light curves. The procedure was repeated
for each of the three color band datasets. In each case, the
resulting light curve indicated evidence for residual emission
during the presumed `off' phase of emission, as can be seen
in Figure 1.\\

It is clearly necessary to determine the duration of the true `off' 
phase of emission, namely that consistent with emission from a constant source.
Perhaps more critically, we want to ensure that this emission is not 
contaminated by the flux associated with the trailing edge of Peak 2 
and the leading edge of Peak 1. In order to do this, we attempted to isolate
that part of the corrected light curve within this `off' phase
region whose phase-average flux is, to first order, consistent with a
constant source of emission. This was done by starting with the largest
phase range in terms of bins defining the traditional `off' region, 
computing the total flux within this range, and then determining 
the idealised average flux level per bin. The deviation over the 
defined phase region of the observed flux levels per bin against 
the averaged flux levels per bin were examined using a $\chi^{2}$ test. 
This process was repeated iteratively, by dropping the test phase 
region window (and hence number of bins), and sweeping this through 
the initially denominated `off' phase region. In this way, at the 
95$\%$ level of confidence, the chosen bin range was 
(0.75 - 0.825) of phase, based on an analysis of the three 
color band light curves. Within this phase region we are 
satisfied that the observed flux is
consistent with emission from a constant source, at this confidence level. We
note that this bin range is marginally smaller than that defined by 
Percival et al. 1993, on analysis of High Speed Photometer data taken of the
Crab pulsar from the Hubble Space Telescope, using a similar analytical
technique.\\

%The resulting Crab removed image was then used to determine the 
%net background flux per color band, and this was then removed from the 
%full cycle light curves in each band. From this, the presumed `off'
%region was isolated, and a basic algorithm and then tested for $\chi^{2}$ 
%deviation away from the average computed background within this
%chosen bin range. This was repeated iteratively, by dropping the bin 
%range and sweeping through that nominated central steady region.
%In this, way, at the 95$\%$ level of confidence, the chosen bin range was
%(0.75 - 0.825) of phase, based on the three colour band light curves.\\           

With this `off' region so defined, the 
corresponding 2-d images were acquired for the three color bands.
Application of the IRAF {\tt allstar} task using the empirically
derived full cycle PSF for each of the $UBV$ images successfuly
removed the faint stellar point source visible in each, and from
this the flux was estimated. In addition, a local PSF was 
constructed per phase-resolved image, and the fitting-and-extraction
process was performed using both local and full-cycle determined
PSFs. This was done for completeness, although the full cycle PSF 
were found to be sufficient and more ideal, being based upon a 
higher S/N source and substantially diminished background noise (in
comparison to the phase-resolved images). This would seem to indicate
that sharp nebular features which might be expected to 
"contaminate" the off pulse 
PSF more than that of the on pulse PSF do not contribute
significantly to these results.\\

The original removal and estimation of the relative fluxes from the 
full cycle $UBV$ datasets yielded a set of reference count rates.
All subsequent flux estimates for specific phase regions were
subsequently normalized to these reference count rates per color band. 
Limited prior observations of several Landolt reference stars in the 
PG0220 field (\cite{land92}) provided calibration magnitudes which indicated
integrated Crab fluxes in agreement with that expected. Using the $UBVR$
ground-based fluxes of Percival et al. 1993  as reference points, 
we thus renormalized
our previously determined fractional fluxes. This reference data was based on
ground based observations of the Crab pulsar made at the 2.1m telescope at
McDonald Observatory in January 1992, and corrected for interstellar
extinction using $E(B-V)$ = 0.51 $\pm$ 0.04 (Savage \& Mathis 1978).\\

Table \ref{table2} details this phase averaged
flux, and in Table \ref{table3} we show the derived fractional
fluxes for that of the unpulsed components as determined by this analysis
in addition to the other light curve components.
In Table \ref{table4} we have reproduced the estimated power-law
parameter $\alpha$ determined via a weighted least-squares analysis
of each individual spectral dataset. We have re-calculated
$\alpha$ for the both the full range \& $UBV$ Percival et al.  dataset
to compare with the other power-law fits. 
We note that one would estimate a change in flux of $\sim$ 0.01 over the
four years between the reference integrated flux and our observations,
following the phenomenologically derived $\dot{L_{V}}$ $\sim$ 0.003 mag/yr
(\cite{pac71}), empirically confirmed most recently by \cite{nasuti97}, which
is within the error bounds quoted.\\

\placetable{table2}
\placetable{table3}
\placetable{table4}

\section{Discussion}

The question of the unpulsed component of emission for the Crab pulsar has
always remained somewhat challenging, as one is confronted with temporal 
problems and the nebular contribution. With this 2-d $MAMA$ data, definitive
flux estimates are attainable for the first time. In Peterson et al. 1978,
(and elsewhere Miller \& Wampler, 1969), the  estimated total unpulsed
emission is compared with the peak intensity - rather a relative area in terms
of phase allocation at our level of temporal resolution - and also with mean
pulsed flux.  Peterson et al. applied rather novel techniques in the image
processing their data obtained via the use of a 6.2ms time resolved Image
Photon Counting System camera. Using an iterative least-squares semi-empirical
based PSF, they determined residuals which when smoothed yielded a background
image which was subtracted from the star field, and the same method estimated
the star intensities.  Peterson et al. did not present errors associated with
their eventual tabulated results. We note the 6.2ms absolute timing
resolution. This is some $\sim$ 20$\%$ of the light curve, and accurate phase
resolution may not have guaranteed accurate continual phase resolution
photometry. Timing errors, accurate phase resolution and estimation
of the total  aperture background are all guaranteed at unprecedented
resolution with our datasets. From the background corrected light curves, we
can determine  the incident flux within the designated `steady' region of
emission, and then compare it directly with both the total pulsar flux and 
pulsed-only flux. These differences, presented in terms of magnitude 
change, are shown in Table \ref{table5}.\\ 

\placetable{table5}

The tabulated data suggests that the original estimates by Peterson et al. 
1978 were optimistic by typically at least a magnitude, but this is
understandable bearing in mind the rather difficult data and analysis they
were working with. There is agreement to some extent with the trend - the
early datasets suggested that there was a greater ratios in the $B$ in
comparison to the $U$ band, yet no error estimates are included. No $V$ data
was analysed at that time. We note that if one was to assume that the unpulsed
emission was restricted to a specfic phase region, and not assumed to exist
for the entire rotation, then whilst the ratios would drop further, they would
still imply a similar spectral form.\\

\begin{figure} 
\plotone{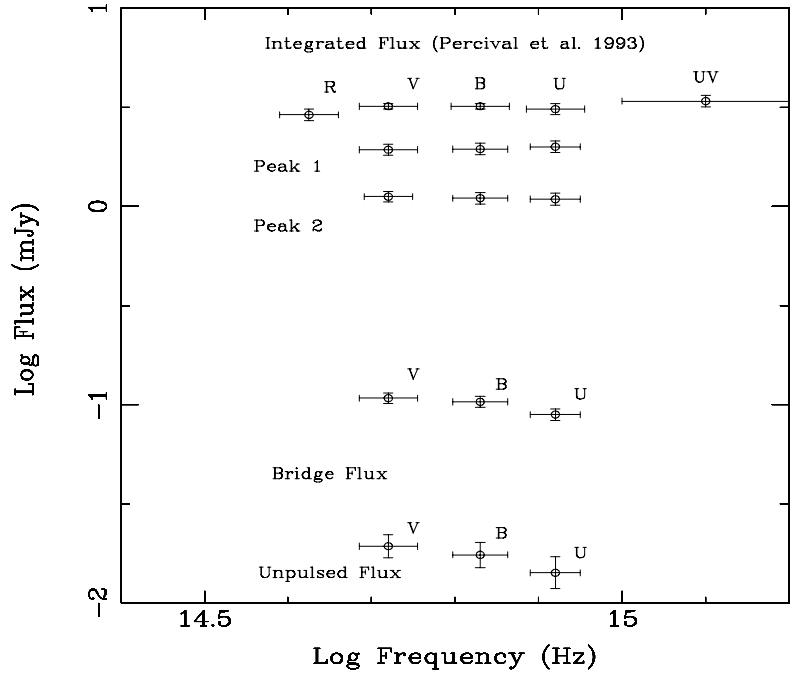}
\caption{Integrated de-extincted flux estimates for the Crab 
from Percival et al. 1993 and the derived flux estimates for 
both the peaks, Bridge and unpulsed component of emission 
based on the TRIFFID datasets.\label{fig2}}
\end{figure}

In Figure \ref{fig2} we reproduce the full Percival et al. (1993)  
derived corrected flux distribution with the unpulsed flux estimates implied
from the tabulated ratios. We have also included the derived flux fractions
for peaks 1 and 2, and the Bridge of emission, which are considered elsewhere
in some detail (Golden et al., 1999). It seems clear that one can
represent the unpulsed emission in spectrally in terms of a steeper power-law
with $\alpha$ $\sim$ -0.60 $\pm$ 0.37 in contrast to the rather flat $\alpha$
$\sim$ 0.11 $\pm$ 0.08 associated with the full integrated emission.\\

\placefigure{fig2}

\section{Conclusion}

The resolved unpulsed flux component, whether within its defined `off' 
region or normalized to the pulsar's full cycle, suggests a power-law form.
% that
%contradicts that expected from a thermal source - we conclude
%that it is not thermal in nature, the anticipated magnitude from a
%$\sim~ 10^{6}$ K neutron star being $\sim$ 30 in V and as such undetectable
%during an exposure of this length. i
There are two options - either the
emission is real and of a nonthermal nature or the emission is false,
a consequence of some form of photocathode or other artifact intrinsic
to the $MAMA$ photon counting detector. This latter would manifest 
timing irregularities 
which
were not evident under analysis. 
Photon timeseries taken from the pulsar
and other stars in the field were tested for deviations from a Poissonian
distribution at varying timescales, and there was no evidence for such
a deviation at the 99\% confidence level, atmospheric variations notwithstanding.
In this, we confirm the earlier work of Smith et al. (1978). 
Consequently we may conclude that the emission is from the pulsar.\\ 

This unpulsed emission has been more commonly observed in the higher
(X-ray \& $\gamma$-ray) regimes, and scrutinised in some detail. 
In X-rays, the unpulsed component is difficult to discern amid the
intense nebular emission. 
Becker \& Aschenbach 1995 attempted to analyse $ROSAT$ HRI 
data, ostensibly to determine limits to the pulsar's thermal emission
during the unpulsed phase - they concluded with a realistic upper limit to
$T_{surface}$ for the Crab's temperature. This does seem to suggest that
in X-rays, the hot Crab and the plerion would dominate the emission.\\  
 
For detected unpulsed $\gamma$-ray emission, the
existing models place the emission either just beyond the magnetosphere
or far out in the plerion, namely the Outer-Gap model of Cheung $\&$ Cheng 
1994, and the pulsar-wind model of De Jager $\&$  Harding 1992. 
Two principle predictive facts concern us regarding these two models;
the first is that the Pulsar-Wind model implies an emission region
large in extent, perhaps up to $\sim$ 20", whereas the Outer-Gap model
requires emission to occur in the immediate vicinity of the pulsar, thus
having a resolution of order $\ll$ 1". Secondly, the Outer-Gap model
implicitly expects a correlation between the pulsed and unpulsed emission,
whether it be temporal or spectral in nature (Cheung $\&$ Cheng 1994).\\ 

Based upon our resolved functional form
for the unpulsed component we can reject the De Jager \& Harding model, as the source is undeniably localised to the
pulsar. The conclusion is that the emission is in
some way magnetospherically related. We cannot accept 
the opposing Cheng \& Cheung model
for a number of reasons. The emission mechanism is
based on the original outer-gap magnetospheric model \cite{chr}, and
in this case is a result of the cross-streaming of two opposing outer-gaps
primary \& secondary photon streams. Here the inner streaming
IR-optical  photons from the far-side gap collide with the primary
$\gamma$-rays \& e$^{\pm}$ pairs from the near-side gap at some distance
($\sim$ 3$R_{LC}$) from the magnetosphere. These interactions result in
isotropically radiated high energy emission, predominately from X-rays (MeV) 
to $\gamma$-rays (Gev -- TeV). Cheung \& Cheng 1994 point out that at the low
(keV - 50 MeV) range, their model predicts flux levels much lower than that
observed, and that other mechanisms not accounted for (they suggest
synchrotron self-Compton mechanisms) must be present. It is clear that any
IR-optical photons that are emitted will be predominately pulsed in nature (as
in the original \cite{chr} ansatz) as the process advocated would be expected
to be preferentially luminous at the higher frequency ranges.\\

There have been other attempts to explain the observed steady
emission (Peterson et al. 1978); they are that

\begin{itemize}
\item the unpulsed component is actually pulsed emission emitted
from points spatially extended in the magnetosphere, and it is
manifested to us following varying time-of-flights and relativistic effects,
\item the pulses could actually possess trailing \& leading edges that
effectively result in fully pulsed emission,
\item the unpulsed component is a result of the reprocessing or reflection
of pulsed emission from material near the pulsar (such as a nebulosity,
localised knot etc.).
\end{itemize}  

The spatially extended hypothesis above is commensurate with the
numerical model framework of Romani and Yadigaroglu 1995, which
requires that emission occurs from such a similar topology
with similar arguments for the resulting formation of the
light curve morphologies. However, this model
was based on a number of questionable assumptions as noted by 
the paper's authors. More critically, Eikenberry \& Fazio 1997 have
unambigously shown evidence for significant intra-color phase differences
between the leading and trailing edges from $\gamma$-rays to the IR,
consistent with a {\it localised} origin. These caveats have made such a
theoretical basis difficult.\\
 
We have already noted the similar spectral forms of both the Bridge and
unpulsed components of emission  as is evident from Table \ref{table4}.
It is our contention that the observed unpulsed component
of emission has its source in a similar electron population/magnetic
field/Lorentz factor environment to the Bridge component. 
The change in power-law exponents from the peaks to the Bridge/unpulsed
component may be as a result of either a change in the emitting
e$^{\pm}$ energy distribution or via modification (due to
scattering or absorption processes) of the emitted photon flux.
In either case this would be consistent with emission occurring
from a region closer to the neutron star within the magnetosphere
particularly if we were to assume a common e$^{\pm}$ energy 
distribution originating above a polar cap, which would be expected
to evolve in this way as the e$^{\pm}$ population streams 
radially along the open field lines.
Ultimately whether both Bridge \& unpulsed emission 
are associated with the main peaks, or whether they are spatially \&
energetically seperate is at this stage unresolved; 
viewing geometry issues may be a major factor. 

We recall from Smith et al. 1988 that the unpulsed component
could be regarded as an extended source of emission, spread
in longitude in proximity to light cylinder. The observed flux would
then be the result of emission from field lines at and beyond the
limit of the polar cone regions, including both the leading and trailing edges
of the cores. 
These field lines would be expected to be affected by
abberation and tend towards a toroidal direction - Smith et al. 1988
note that the position angle of the unpulsed (and indeed Bridge component)
is similar to that of the mean of the peaks, namely $130^{o}$.
Thus these unusual polarisation effects noted by Smith et al. 1988
and others indicate similar behaviour for both Bridge \& unpulsed
regions, perhaps substantiating a belief that they are 
in some way phenomenologically linked. Another hypothesis
is that the observed unpulsed component represents 
some fraction of the original synchrotron emitting photons
scattered by the local $e^{\pm}$ particle density along various
path lengths within the magnetosphere, resulting in an
apparently isotropically radiated emission component.
However, one might expect an essentially randomised polarisation
nature to these scattered photons, which is not reflected in
previous phase-resolved polarimatory.\\

Clearly, we require further unpulsed estimates in the $UV$ and 
$R-IR$ wavebands in order to characterise the manner in which this emission
component correlates with that of the dominant pulsed emission - most
interestingly in the vicinity of $\sim$ $10^{14}$ Hz, where an apparent
rollover inconsistent with conventional synchrotron self-absorption is
apparent. Such estimates would provide constraints to the existing
power-law fits, consolidating our contention that the unpulsed component
of emission is steeper than that for the integrated spectral index.
Perhaps of even greater urgency would be the definitive acquistion
of polarimetric photometry of the unpulsed component with the nebular content
removed so as to finally assess a possible link between it and the Bridge
of emission. Such future work could provide yet more critical empirical
constraints to the nascent field of numerical magnetospheric 
optical emission models.\\

Acknowledgements

The authors wish to thank R. Butler for assistance with the photometric
analysis and M. Cullum for provision of the ESO MAMA detector.
The support of Enterprise Ireland, the Irish Research and Development 
agency, is gratefully acknowledged. This work was supported by the 
Russian Foundation of Fundamental Research, the Russian Ministry of 
Science and Technical Politics, and the Science-Educational 
Centre "Cosmion", and under INTAS Grant No. 96-0542.

\clearpage

\clearpage

\clearpage

\begin{deluxetable}{cccccc}
\footnotesize
\tablecaption{Summary of Observations January 1996 {\bf BTA}\label{table1}}
\tablewidth{0pt} 
\tablehead{
\colhead{Dataset}  &   \colhead{Date}   &  \colhead{UTC}    & 
\colhead{Duration (s)} & \colhead{Filter}  &  \colhead{Seeing (")} 
} 
\startdata 
96zd3.0.0  & 12/1/96 &  16:38:51 & 155  &   V   &   2"   \\ 
96zj5.0.0  & 13/1/96 &  16:33:50 & 334  &   U   &  1.6" \\ 
96zj7.0.0  & 13/1/96 &  16:53:36 & 297  &   V   &  1.3" \\ 
96zo5.0.0  & 14/1/96 &  16:07:36 & 617  &   U   &  1.9" \\ 
96zo6.0.0  & 14/1/96 &  16:21:41 & 584  &   V   &  1.6" \\ 
96zs1.0.0  & 15/1/96 &  17:07:34 & 1502 &   U   &  1.5" \\ 
96zs2.0.0  & 15/1/96 &  17:38:20 & 1177 &   V   &  1.3" \\
96zt2.0.0  & 15/1/96 &  18:48:40 & 170  &   V   &  1.4" \\
96zv2.0.0  & 16/1/96 &  16:37:01 & 1468 &   V   &  1.8" \\
96zw2.0.0  & 16/1/96 &  17:38:01 & 1811 &   B   &  1.6" \\
96zw3.0.0  & 16/1/96 &  18:09:16 &  920 &   V   &  1.4" \\
96zw4.0.0  & 16/1/96 &  18:44:09 & 1415 &   B   &  1.4" \\
96zx1.0.0  & 16/1/96 &  19:08:51 &  140 &   V   &  1.4" \\
96zx2.0.0  & 16/1/96 &  21:15:04 &   49 &   B   &  1.7" \\
96zx3.0.0  & 16/1/96 &  21:20:22 & 1212 &   B   &  1.7" \\
96zx9.0.0  & 17/1/96 &  16:02:33 &  337 &   U   &  1.8" \\
96zx10.0.0 & 17/1/96 &  16:09:51 &   47 &   U   &  1.8" \\
96zx11.0.0 & 17/1/96 &  16:15:32 & 2709 &   U   &  1.9" \\
96zy1.0.0  & 17/1/96 &  17:02:42 & 1848 &   V   &  1.6" \\
96zy3.0.0  & 17/1/96 &  17:39:08 & 1263 &   B   &  1.7" \\
\enddata
\end{deluxetable}

\clearpage

\begin{deluxetable}{cccc}
\footnotesize
\tablecaption{Integrated Flux from the Crab Pulsar (Percival et al.
1993). The third column represents the fractional transmission
as a function of the extinction per specific waveband.\label{table2}}
\tablewidth{0pt}
\tablehead{
\colhead{Band}  &   \colhead{Raw Flux Density}   & \colhead{Extinction}    &
\colhead{De-extincted Flux Density}\\
\colhead{}  &   \colhead{mJy}   & \colhead{Frac. Trans.}    &
\colhead{mJy}
}
\startdata
UV &  0.11 $\pm$ 0.02 & 0.031 $\pm$ 0.002 & 3.4 $\pm$ 0.25   \\
U  &  0.31 $\pm$ 0.02 & 0.101 $\pm$ 0.006 & 3.1 $\pm$ 0.2    \\
B  &  0.47 $\pm$ 0.02 & 0.145 $\pm$ 0.009 & 3.2 $\pm$ 0.2    \\
V  &  0.73 $\pm$ 0.04 & 0.227 $\pm$ 0.014 & 3.2 $\pm$ 0.2    \\
R  &  0.90 $\pm$ 0.05 & 0.313 $\pm$ 0.019 & 2.9 $\pm$ 0.2    \\
\enddata
\end{deluxetable}

\clearpage

\begin{deluxetable}{llll}
\tablecaption{Fractional Flux derived from Photometric Analysis.\label{table3}}
   \tablewidth{0pt} 
\tablehead{   
\colhead{Parameters}    &    & \colhead{Waveband}  &     \\
 &   \colhead{U  (mJy)}     &  \colhead{B  (mJy)} &  \colhead{V (mJy)}
} 
\startdata
Peak 1                  & 2.00   $\pm$ 0.13  & 1.95  $\pm$ 0.12   & 1.93  $\pm$ 0.12 \\
Peak 2                  & 1.09   $\pm$ 0.07  & 1.1   $\pm$ 0.07   & 1.12 $\pm$ 0.07 \\
Bridge                  & 0.087  $\pm$ 0.006 & 0.103 $\pm$ 0.006  & 0.108 $\pm$ 0.007 \\
Off Phase               & 0.014  $\pm$ 0.002 & 0.017 $\pm$ 0.002  & 0.019 $\pm$ 0.002 \\
\enddata
\end{deluxetable}

\clearpage

\begin{deluxetable}{cc}
\tablecaption{Estimated Spectral Power-Laws from Photometric
Analysis\label{table4}}
\tablewidth{0pt}
\tablehead{
\colhead{Dataset}  &   \colhead{Power-Law $\propto$ $\nu^{\alpha}$}          
}
\startdata
Integrated UV/U/B/V/R   & 0.11  $\pm$ 0.09     \\
Integrated U/B/V        & -0.07 $\pm$ 0.18     \\
Peak 1                  &  0.07 $\pm$ 0.19   \\
Peak 2                  & -0.06 $\pm$ 0.19   \\
Bridge                  & -0.44 $\pm$ 0.19    \\
Off                     & -0.60 $\pm$ 0.37      \\
\enddata
\end{deluxetable}

\clearpage

\begin{deluxetable}{ccccc}
\tablecaption{Ratios of pulsed/unpulsed emission in magnitudes for $U$, $B$ \& $V$. Here, total
emission corresponds to the full integrated emission from the pulsar. Full cycle
corresponds to the unpulsed emission over 0.075 phase normalised to one
cycle.\label{table5}}
\tablewidth{0pt}
\tablehead{
Ratio        &   U  &  B  &   V    
}
\startdata
\null & ${\bf This ~Work}$ & \null & \null & \\
\hline
total emission/(0.075 phase) &  5.8   $\pm$ 0.2       & 5.6 $\pm$ 0.1  & 5.6  $\pm$
0.1 \\
total emission/(full cycle)   &  3.0  $\pm$ 0.2       & 2.8 $\pm$ 0.1  & 2.7  $\pm$
0.1 \\
\hline
 \null & ${\bf Peterson ~et ~al.}$ 1978 & \null & \null \\
\hline
total emission/(full cycle)  & 1.5 & 1.6 & n/a \\
\hline
\enddata
\end{deluxetable}

\end{document}